\documentclass[]{spie}  
\usepackage[]{graphicx}
\usepackage{amssymb,amsmath}
\usepackage{subfig}

\newcommand{\myFigureWidth}{0.63}
\newcommand{\myHalfFigureWidth}{0.45}

\title{Status of SuperSpec: A Broadband, On-Chip Millimeter-Wave Spectrometer} 

\author{S. Hailey-Dunsheath\supit{a}, E. Shirokoff\supit{b}, P. S. Barry\supit{c}, C. M. Bradford\supit{d}, G. Chattopadhyay\supit{d}, P. Day\supit{d}, S. Doyle\supit{c}, M. Hollister\supit{a}, A. Kovacs\supit{e}, H. G. LeDuc\supit{d}, P. Mauskopf\supit{c,g}, C. M. McKenney\supit{a}, R. Monroe\supit{a}, R. O'Brient\supit{a}, S. Padin\supit{a}, T. Reck\supit{d}, L. Swenson\supit{a}, C. E. Tucker\supit{c}, and J. Zmuidzinas\supit{a,d}
\skiplinehalf
\supit{a}California Institute of Technology, Pasadena, CA, U.S.A. \\
\supit{b}University of Chicago, Chicago, IL, U.S.A \\
\supit{c}School of Physics \& Astronomy, Cardiff University, Cardiff, U.K. \\
\supit{d}Jet Propulsion Laboratory, Pasadena, CA, U.S.A. \\
\supit{e}University of Minnesota, Twin Cities, MN, U.S.A. \\
\supit{f}Delft University of Technology, Delft, Netherlands \\
\supit{g}Arizona State University, Tempe, AZ, U.S.A
}

\authorinfo{Direct correspondence to S. Hailey-Dunsheath: haileyds@caltech.edu}

\begin{document} 
  \maketitle 

\begin{abstract}

SuperSpec is a novel on-chip spectrometer we are developing for multi-object, moderate resolution ($R = 100 - 500$), large bandwidth ($\sim$$1.65$:$1$) submillimeter and millimeter survey spectroscopy of high-redshift galaxies. The spectrometer employs a filter bank architecture, and consists of a series of half-wave resonators formed by lithographically-patterned superconducting transmission lines. The signal power admitted by each resonator is detected by a lumped element titanium nitride (TiN) kinetic inductance detector (KID) operating at $100-200$ MHz. We have tested a new prototype device that is more sensitive than previous devices, and easier to fabricate. We present a characterization of a representative $R=282$ channel at $f = 236$ GHz, including measurements of the spectrometer detection efficiency, the detector responsivity over a large range of optical loading, and the full system optical efficiency. We outline future improvements to the current system that we expect will enable construction of a photon-noise-limited $R=100$ filter bank, appropriate for a line intensity mapping experiment targeting the [CII] 158 $\mu$m transition during the Epoch of Reionization.

\end{abstract}



\section{INTRODUCTION} \label{introduction}
Over the history of the Universe, much of the radiation generated by stars and black holes at ultraviolet through near-infrared (IR) wavelengths has been absorbed by dust, and reradiated in the far-IR. This process produced the cosmic far-IR background (CFIRB)\cite{Fixsen1998}, which has an integrated intensity comparable to that of the cosmic optical background, demonstrating that obscured star formation and black hole growth were important in the history of galaxies. Understanding the nature of this obscured activity is a key goal of galaxy evolution studies. 

Surveys at far-IR through millimeter wavelengths are revealing the sources that created this background\cite{Berta2010,Vieira2013,Geach2013SCUBA2}. Most are likely at redshifts $z < 3$, but selection based on far-IR colors can identify objects as high as $z \gtrsim 6$\cite{Dowell2014}. Spectroscopic follow-up of these sources at optical wavelengths is challenging, due to the significant extinction that may be present in the galaxy. A more reliable approach is to target atomic and molecular line emission emitted in the rest-frame far-IR through millimeter wavelengths, where dust extinction is minimized. Detection of these lines in high-redshift systems has been achieved using both direct-detection grating spectrometers\cite{Stacey2010CII,Riechers2013Nature} and heterodyne receivers\cite{Weiss2009,Walter2012Nature}. However, due to the large size of grating spectrometers, and the limited bandwidth and field of view of millimeter-wave interferometers, a new approach is required to efficiently conduct broadband spectroscopy of large samples.

Development of a millimeter-wave multi-object spectrometer will also enable the construction of line intensity mapping experiments for studies of large-scale structure. In particular, the [CII] 158 $\mu$m fine-structure line is redshifted into the 1 millimeter telluric window for $z = 5-9$, providing a probe of the Epoch of Reionization (EoR)\cite{Gong2012}. Crucially, the galaxies producing the bulk of the reionizing photons during this epoch are expected to be small, low luminosity systems, and difficult to detect individually\cite{Salvaterra2011}. However, the power spectrum obtained from a 3-D spatial/spectral map of the [CII] line traces the aggregate emission from all sources at this epoch, providing a measure of the galaxy clustering, and constraints on the galaxy luminosity function. A cross-correlation of the [CII] map with a map of the HI 21-cm line emission additionally probes the detailed physics of reionization, including the evolution of ionizing bubble sizes and the mean ionization fraction\cite{Gong2012}.

To address these needs we are developing SuperSpec, a broadband, on-chip millimeter-wave spectrometer. Its small size, large spectral bandwidth, and highly multiplexed readout will enable construction of powerful multi-beam spectrometers for high-redshift observations. We are currently developing this technology with $R\gtrsim250$ spectrometers operating in the $190-310$ GHz band, with the aim of deploying a future wide-band survey spectrometer for the 25 meter CCAT telescope. The proposed instrument will employ tens of thousands of detectors in tens of beams covering the $190-510$ GHz band, and will be optimized for measuring the bright atomic fine-structure and molecular rotational lines from interstellar gas in galaxies at $z\approx3-9$ (C. M. Bradford et al., these proceedings). A lower resolution ($R\sim100$) spectrometer is also a candidate technology for an EoR line intensity mapping experiment, such as the proposed TIME (A. Crites et al., these proceedings). In this paper we describe recent progress in the development of this technology.

\section{SuperSpec Concept and Circuit Design} \label{concept}

The SuperSpec design follows the concept of a filter bank spectrometer presented by Kov{\'a}cs et al.\cite{Kovacs2012SPIE} (see also R. O'Brient et al., these proceedings). Radiation propagating down a transmission line encounters a series of tuned resonant filters, each of which consists of a section of transmission line of length $\lambda_i/2$, where $\lambda_i$ is the resonant wavelength of channel $i$. These half-wave resonators are coupled to the feedline and to power detectors with adjustable coupling strengths, described by quality factors $Q_\mathrm{feed}$ and $Q_\mathrm{det}$, respectively. Accounting for additional sources of dissipation in the circuit with a coupling factor $Q_\mathrm{loss}$, the spectrometer resolving power $R$ is equal to the net filter quality factor $Q_\mathrm{filt}$, and is given by:

\begin{subequations}
\begin{align}
\frac{1}{R} &= \frac{1}{Q_\mathrm{filt}} = \frac{1}{Q_\mathrm{feed}} + \frac{1}{Q_\mathrm{det}} + \frac{1}{Q_\mathrm{loss}}.
\end{align}
\end{subequations}

\noindent The filter bank is formed by arranging a series of channels monotonically decreasing in frequency, with a spacing between channels equal to an odd multiple of $\lambda_i/4$. The oversampling factor $\Sigma$ is defined as the ratio of channel width to channel spacing. For a single isolated channel the maximum detection efficiency occurs when $Q_\mathrm{feed} = Q_\mathrm{det}$, and is 50\% on-resonance for the case of no additional loss. Increasing the oversampling factor $\Sigma$ such that the passbands of adjacent channels overlap will increase the total in-band absorption efficiency. For example, a simulated 540 channel, $R = 600$ filter bank covering $195-310$ with $\Sigma = 1.95$ has greater than 80\% absorption in band.

\begin{figure}
\begin{center}
\includegraphics[
  width=0.9\linewidth,
  keepaspectratio]{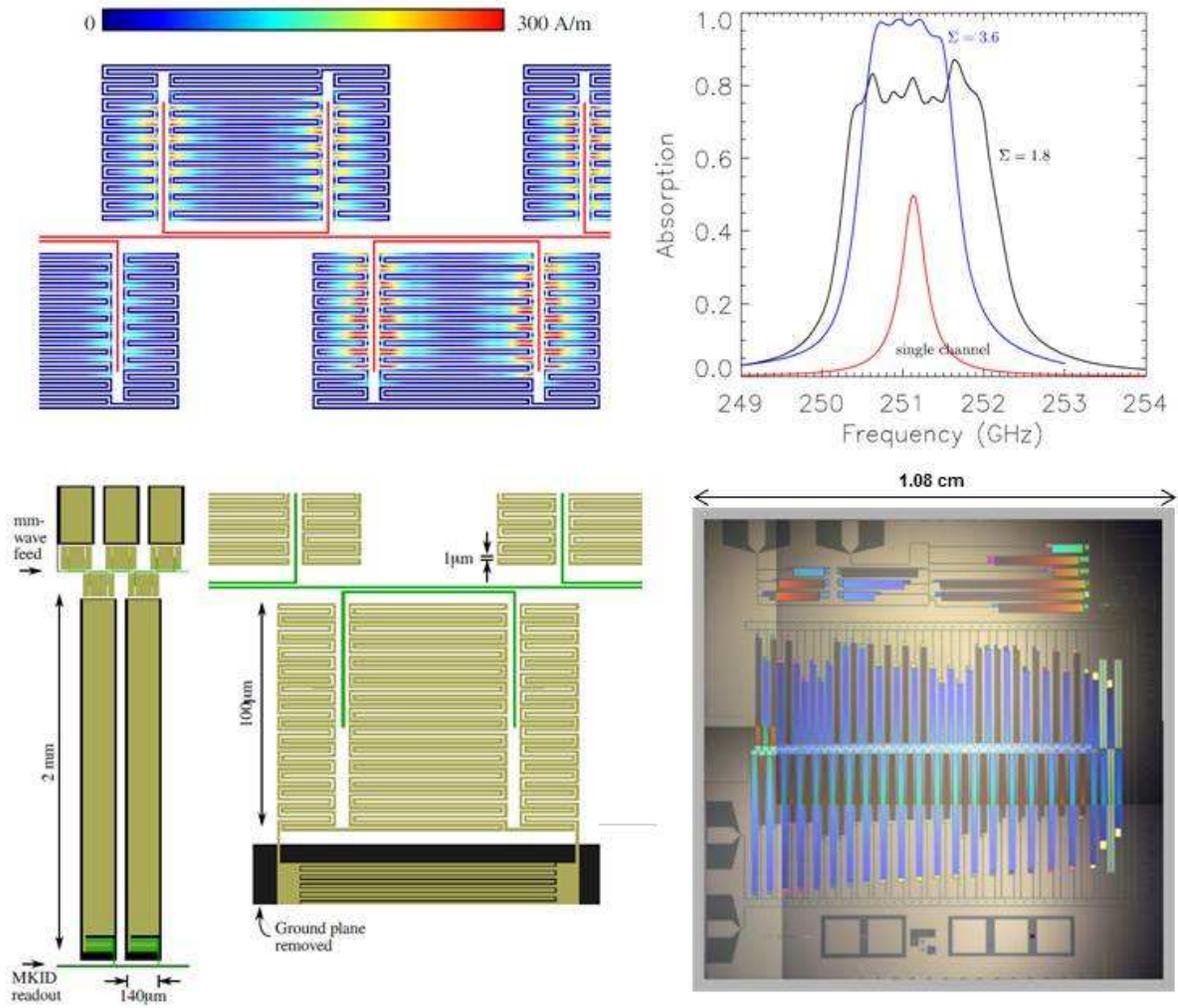}
\end{center}
\caption{(\textbf{top left}) Simulation showing the time-averaged magnitude of the current in a subsection of an 8-channel Gen 1 SuperSpec filter bank, when driven at a single frequency. Millimeter-wave radiation incident from the left along the central feedline couples to U-shaped half-wave resonators, which in turn excite currents in the TiN meanders. (\textbf{top right}) Total power absorbed for the 8-channel filter bank with two different oversampling factors, compared with the response of a single isolated channel. (\textbf{bottom center}) Millimeter-wave portion of a single channel, showing the Nb microstrip feedline and half-wave resonator (\textit{green}), and the TiN meander and top portion of the TiN IDC (\textit{amber}). (\textbf{bottom left}) Wider view showing several nearby channels. Below the KID IDC, a second, smaller IDC couples the KID to the microwave readout line. (\textbf{bottom right}) Image of a dark Gen 1 chip. The 81 channel circuit is visible in the middle, with test structures on the top and bottom of the chip. The waveguide probe was not patterned in this chip.}
\label{concept_figure}
\end{figure}

The SuperSpec concept is implemented with thin-film superconducting circuits (Figure~\ref{concept_figure}), and described more fully elsewhere\cite{Shirokoff2012SPIE, Barry2012SPIE}. Free space radiation is coupled into the feedline from a broadband antenna. Both the feedline and the resonators are inverted microstrip, consisting of Nb traces on Si substrate beneath an amorphous silicon-nitride dielectric and Nb ground plane. The signal power admitted by each resonator is dissipated in a segment of lossy meander formed from titanium nitride (TiN)\cite{LeDuc2010}. Radiation at frequencies above the superconducting gap in the TiN film ($f\sim73$ GHz for $T_c\sim1$ K) breaks Cooper pairs and generates quasiparticles, resulting in a perturbation of the complex impedance.

The TiN meander is connected in parallel to an interdigitated capacitor (IDC) made from the same TiN film to form a lumped element kinetic inductance detector (KID). Perturbations to the complex impedance of the meander translate into changes in the dissipation and resonant frequency of the KIDs, which operate in the $100-200$ MHz range. Each KID is coupled to a coplanar waveguide readout feedline (CPW) by a small coupling capacitor, formed by TiN on the KID side, and Nb on the readout side. The KIDs are coupling-$Q$ limited with $Q_r > 10^5$, where the low readout frequency and high $Q_r$ are designed to minimize the readout bandwidth per channel, thereby maximizing the multiplexing density. The chip is cooled by a $^3$He sorption refrigerator to 220 mK, sufficiently below the TiN $T_c$ to keep the thermal quasiparticle density low, while also as high as possible to minimize two-level system (TLS) noise\cite{Zmuidzinas2012}.

\section{First Generation Prototype} \label{firstgen}

We previously reported on the design and characterization of a first generation (Gen 1) SuperSpec prototype\cite{HaileyDunsheath2014LTD,Shirokoff2014LTD}, and here we briefly review these results. The optical coupling into the Gen 1 chip was provided by a direct-drilled, profiled horn\cite{Leech2011}. The circular waveguide output of the horn is converted to a single-mode oval waveguide and fed with a planar probe, fabricated from the 25 $\mu$m thick device layer of a SOI wafer that supports the spectrometer chip. The probe couples through a CPW transition segment to the microstrip that forms the spectrometer feedline.

The Gen 1 filter bank included 73 spectral channels with a range of $Q_\mathrm{feed}$ and $Q_\mathrm{det}$ values, as well as 8 broadband detectors included for diagnostic purposes. In tests of several dies, the median yield for operable KIDs was 78 of the 81 channels, with no array-disabling critical flaws, while the readout $Q_r$ values were consistently close to the design value of $2\times10^5$. The $T_c$ of the best-characterized device was $\approx$$2$ K.

\begin{figure}%
    \centering
    \subfloat[]{{\includegraphics[width=\myHalfFigureWidth\linewidth,keepaspectratio]{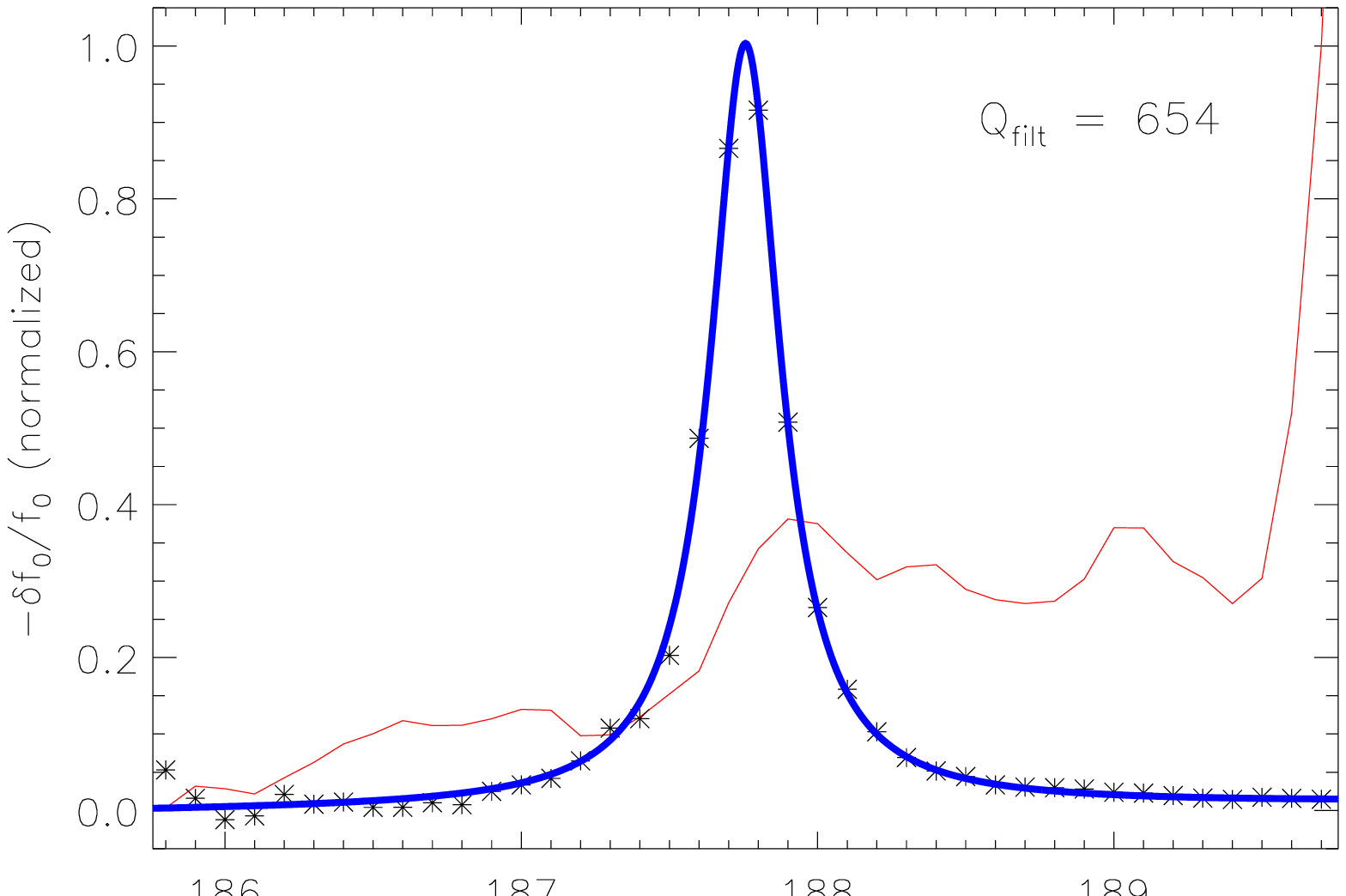} }}%
    \qquad
    \subfloat[]{{\includegraphics[width=\myHalfFigureWidth\linewidth,keepaspectratio]{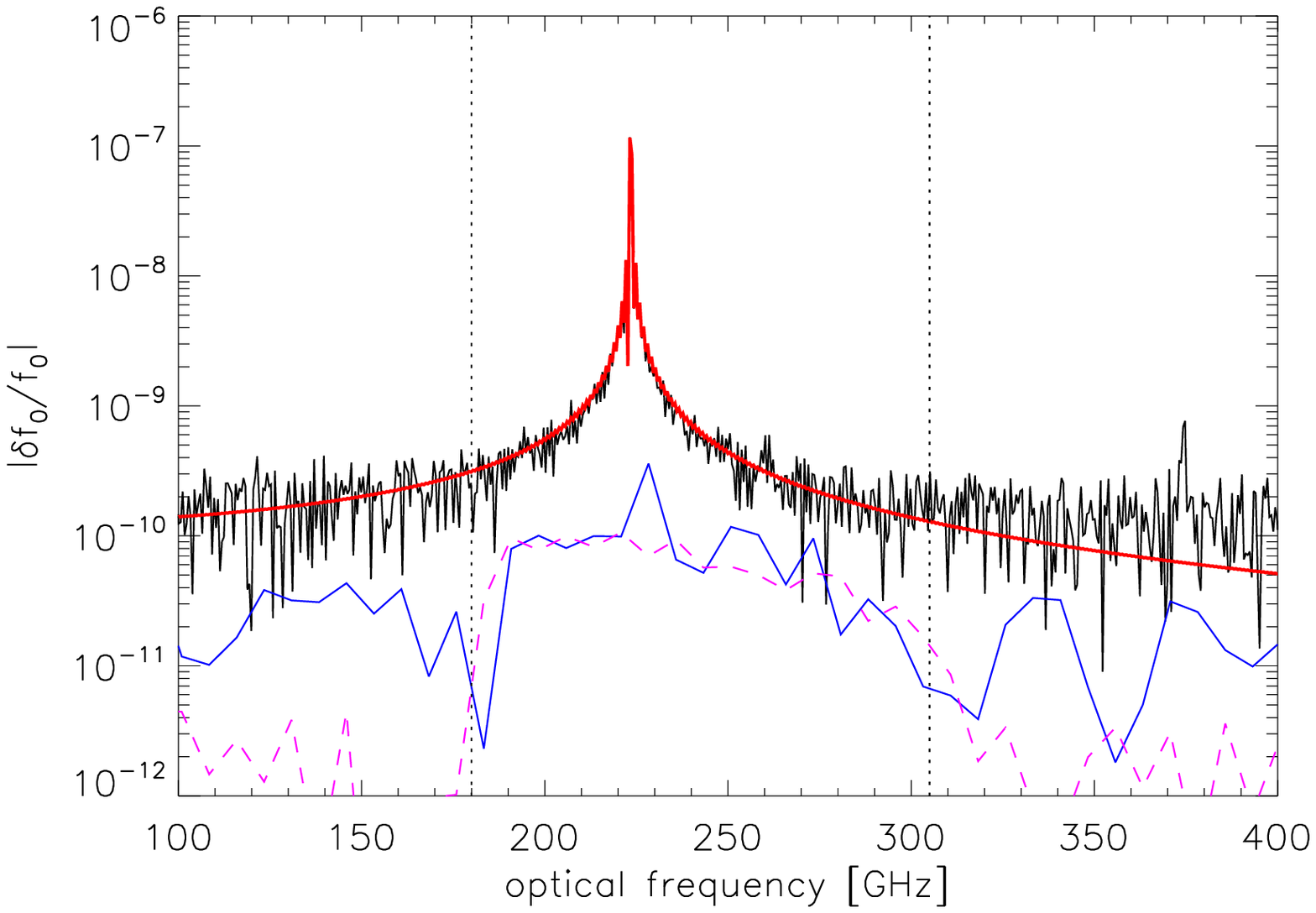} }}%
    \caption{(\textbf{a}) Spectral profile of a Gen 1 $Q_\mathrm{filt} = 654$ channel measured with a swept coherent source (\textit{black points}) and a Lorentzian fit (\textit{thick blue line}). Data points show the measured response normalized by the response of a broadband KID (\textit{thin red line}). (\textbf{b}) FTS measurement of the spectral profile of a typical channel (\textit{thin black}). Overplotted are a Lorentzian profile fit (\textit{thin red}), the residuals binned (\textit{thin blue}), the scaled profile of a broadband KID (\textit{dashed magenta}), and the expected passband defined by feed horn, waveguide probe, and metal-mesh filters (\textit{dotted black vertical lines}). Outside of the narrow channel profile, the response is $\sim$$0.1\%$ the peak value for most channels, and closely matches the profile measured on the broadband KID.}%
    \label{gen1_spectral_profile}%
\end{figure}

A swept coherent source was used to characterize the profiles of the spectral channels (Figure~\ref{gen1_spectral_profile}). The measured linewidths for channels with targeted widths of $Q_\mathrm{filt} \lesssim 400$ were close to the designed values, while the quality factors for higher-$Q$ channels were worse than designed. A comparison of the measured and designed $Q_\mathrm{filt}$ for a large sample of channels suggested the presence of a loss mechanism characterized by $Q_\mathrm{loss}\approx 1440$, likely the silicon-nitride dielectric layer in the microstrip. Measurements using a Fourier Transform Spectrometer indicated high spectral purity, with a typical out of band response $\sim$$30$ dB below the peak response (Figure~\ref{gen1_spectral_profile}).

The measured fractional frequency noise in these devices was typically $(1-2)\times10^{-17}$ Hz$^{-1}$ at a demodulated frequency of 1 Hz, and was consistent with TLS noise. The response of the spectral channels to a beam-filling thermal source placed outside of the cryostat was lower than expected, based on comparison to Mattis-Bardeen theory or to scaled measurements from detectors in the MAKO camera\cite{McKenney2012}. The relative contributions of poor optical coupling and low intrinsic TiN responsivity in generating this low response were not immediately clear.  

\section{Second Generation Prototype} \label{secondgen}

\subsection{Filter Bank Design}

The results from the Gen 1 chip characterization confirmed the basic functionality of the filter bank, but also demonstrated a need for improved sensitivity. In the second generation (Gen 2) design we made several changes to the chip in order to improve the detector response, reduce TLS noise, enable better characterization of the spectrometer efficiency, and simplify fabrication of prototypes.

\begin{figure}
\begin{center}
\includegraphics[%
  width=1.0\linewidth,
  keepaspectratio]{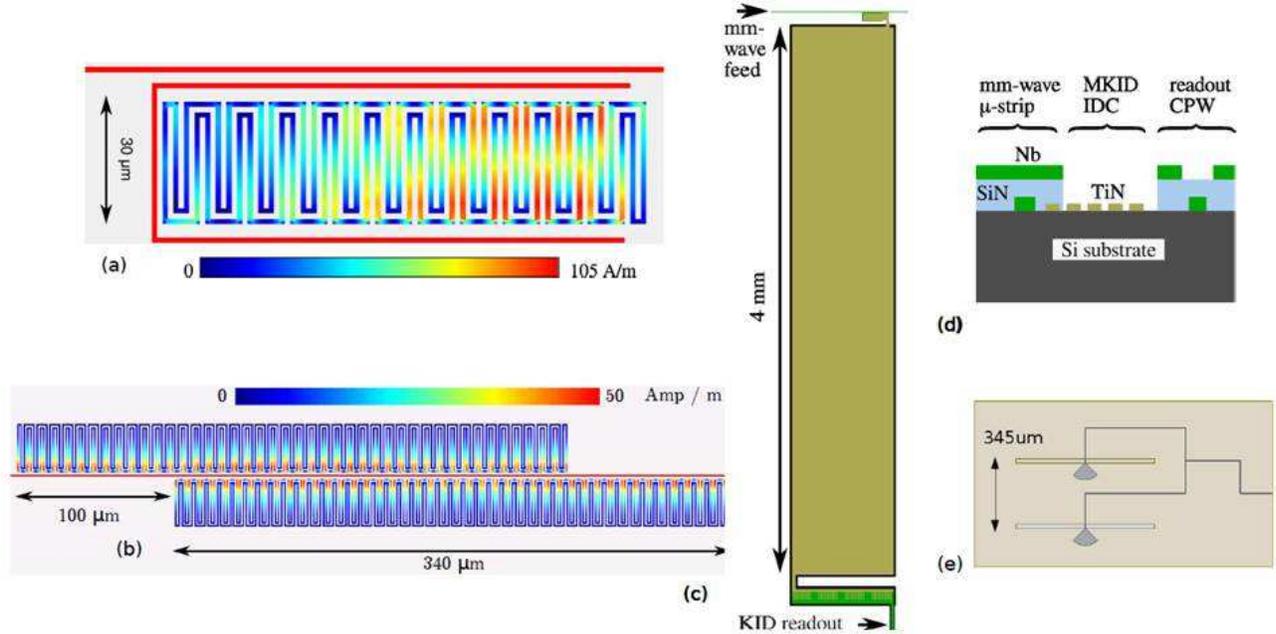}
\end{center}
\caption{Gen 2 device design. (\textbf{a}) Filter bank coupled KID inductor. Simulated time-average millimeter-wave current distribution in the TiN meander is shown, with Nb feedline and U-shaped half-wave resonator off-scale in solid red. (\textbf{b}) A pair of broadband channels arranged on either side of the millimeter-wave feedline. (\textbf{c}) Typical dimensions of one spectral channel. Regions with black outline are where dielectric and ground plane have been removed, leaving only TiN and Nb on bare silicon. (\textbf{d}) Cross-section of the same region. (\textbf{e}) Planar antenna features, showing microstrip (\textit{gray}) and outlined slots in the ground plane.}
\label{fig3}
\end{figure}

The primary change was to reduce the TiN inductor volume by a factor of 4.4, by shrinking the length of the meander, while keeping the previous trace width and film thickness (Figure~\ref{fig3}). A fixed amount of signal power coupling in to the smaller volume meander will produce a larger change to the quasiparticle density, and generate a correspondingly larger response. This was accompanied by a $90^\circ$ rotation of the resonator with respect to the feedline. This rotated geometry increases the coupling between the resonator and feedline, increasing the size of the gap and thus relaxing fabrication tolerances for a given value of $Q_\mathrm{feed}$. To partially compensate for the smaller inductance, the largest capacitor was increased by a factor of 2.4. The lowest readout frequency increased by 35\%. The measurements presented below were obtained using a chip with a TiN $T_c\approx1.65$ K.

The new KID inductor geometry includes two additional enhancements. First, the distribution of millimeter-wave current in the TiN meander, and thus deposited power, is more uniform. The responsivity of the TiN is expected to decrease with increasing local power densities\cite{Zmuidzinas2012}. Thus, non-uniformities in power dissipation on scales larger than the quasiparticle diffusion length reduce the total response compared to what one would predict if the same power were distributed uniformly in the inductor. (If on the other hand the responsivity is independent of power density under our operating conditions, as we report in section~\ref{response_v_loading}, then a more uniform current distribution should not offer any improvement in response.) Second, in the new design the inductor is arranged such that neighboring segments of meander are equidistant from the capacitor, minimizing inductive coupling to the environment.

The IDC feature spacing was increased in order to reduce TLS noise.  The Gen 1 chip had IDCs with either $1 \mu$m line widths and $1 \mu$m gaps between fingers, or $2\mu$m feature widths and gaps. For the Gen 2 design we adopted $1\mu$m finger widths and $3\mu$m spacing. To accommodate the increased physical size of the capacitors, the spacing between filter bank channels on the feedline was increased from $\lambda/4$ to $3\lambda/4$. 


The Gen 2 filter bank contains 10 spectral channels, well isolated in optical and readout frequency space. In addition, the Gen 2 device also includes six broadband absorbers, two placed between the antenna and the filter bank, two placed at the end of the filter bank, and two incorporated into the lossy terminator at the end of the line. These absorbers consist of a meander of TiN placed $1.5\mu$m from the millimeter-wave feedline and connected to an IDC.  Proximity coupling deposits approximately 1\% of the total power on the feedline in each broadband absorber meander in a wide spectral band. Each broadband absorber couples to a region of feedline longer than a wavelength, and pairs of absorbers are offset by approximately 1/4 wavelength, providing robust measurement of power on the feedline even in the presence of standing waves.

To simplify the optical train and prototype fabrication, the probe-coupled feed horn used in the Gen 1 design was replaced by a dual-slot antenna. This antenna provides less bandwidth, but has a history of demonstrated high efficiency, and does not require the complicated backside processing needed for the SOI probe.  We glue a 1.0 cm diameter hyperhemispherical silicon lens on the back of the wafer to produce a Gaussian beam with $\sim$27dB directivity\cite{Filipovic1993ITMTT}.

An image of a Gen 2 chip is shown in Figure 4. After adhering the lens to the back of the wafer, we mount the chip in a light-tight copper box. The lens pokes through a slightly oversized hole, with carbon-loaded Stycast around the hole to make the immediate environment black. A one inch diameter metal cylinder with a rough blackened interior surface supports a metal-mesh low pass filter and prevents reflected stray light at glancing angles from reaching the silicon. The KIDs and microwave readout features are placed offset from the lens footprint, in order to further minimize direct stimulation by stray light.





\begin{figure}
\begin{center}
\includegraphics[
  width=\myFigureWidth\linewidth,
  keepaspectratio]{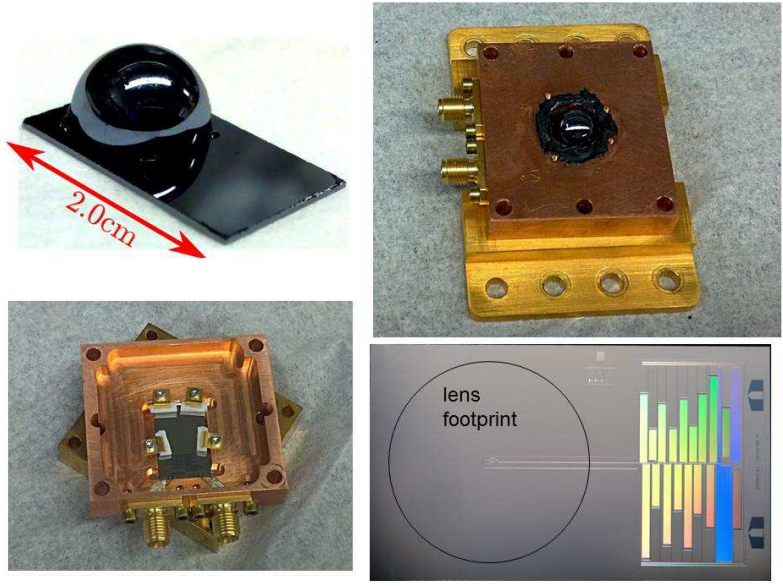}
\end{center}
\caption{Gen 2 die and packaging. (\textbf{top left}) A Gen 2 prototype with mounted lens; devices are on the bottom of the die. (\textbf{top right}) The optical test package, with the lens of a mounted die visible in the center.  An additional blackened cylinder and a metal mesh low-pass filter (not shown) attach to the top surface. (\textbf{bottom left}) The interior of the optical housing with a device in place. (\textbf{bottom right}) The device layout and lens position.  The millimeter-wave feedline extends from the twin slot antenna at the center of the lens footprint horizontally to the right, past a series of filter bank coupled channels on alternating sides of the feedline.}
\label{gen2_image}
\end{figure}

\subsection{Spectrometer Efficiency} \label{spectrometer_efficiency}

The Gen 2 prototype device was tested in an optical cryostat cooled by a commercial pulse tube cooler and a $^3$He sorption refrigerator, providing a 220 mK base temperature. A series of plastic filters and low-pass metal mesh filters in the cryostat provide IR-blocking and band definition. We probe the optical response of the spectrometer using a commercial amplifier/multiplier chain, which provides a $\times18$ multiplication of an RF tone, and an output tunable over $220-330$ GHz\footnote{www.vadiodes.com}. This local oscillator (LO) chain is coupled to a feed horn, and radiates directly into the cryostat. Frequency sweeps of a spectral channel and pairs of broadband absorbers located before and after the filter bank are used to fully characterize the spectral response. Here we present the characterization of a representative moderate resolution ($R\sim250$) channel centered at $f = 236$ GHz.

\begin{figure}
\begin{center}
\includegraphics[
  width=0.95\linewidth,
  keepaspectratio]{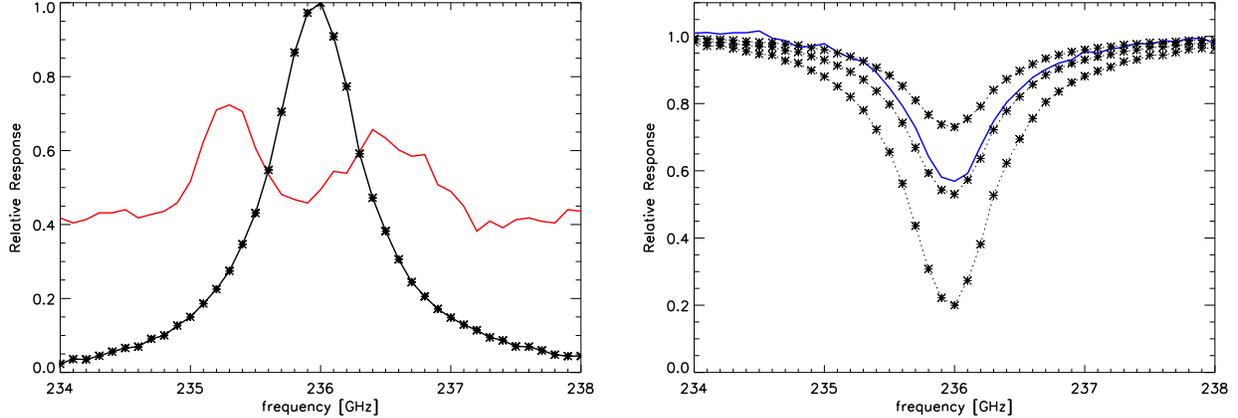}  
\end{center}
\caption{(\textbf{left}) Spectral profile of an $R=282$ channel (\textit{black}), after normalizing by the summed response of two broadband KIDs positioned prior to the filter bank (\textit{red}). (\textbf{right}) Ratio of the summed response of two broadband KIDs positioned after the filter bank to the summed response of two broadband KIDs positioned at the start of the filter bank, normalized to an off-resonance value of unity (\textit{blue}). Also shown are models obtained by inverting the normalized spectral response shown in the left panel, with on-resonance minima obtained from theoretical calculations for $Q_\mathrm{filt}/Q_\mathrm{feed} = 0.15$, 0.25, and 0.50, moving from top to bottom (\textit{black}).}
\label{sockout_figure}
\end{figure}

The results of an LO sweep of this channel are shown in Figure~\ref{sockout_figure}. For each optical frequency we use a VNA sweep to measure the complex transmission $S_{21}(f)$ of the spectrometer and broadband KIDs, and for each KID estimate the resonant frequency. The left panel of Figure~\ref{sockout_figure} shows the frequency shift of the spectral channel normalized by the average frequency response of the pair of broadband KIDs at the front of the filter bank. This ratio provides a measurement of the normalized spectral response function of the spectrometer channel, which is well characterized by a Lorentzian with $Q_\mathrm{filt} = 282$.


The right panel of Figure~\ref{sockout_figure} shows the mean frequency response of the pair of broadband KIDs at the end of the filter bank divided by the mean response of the broadband KIDs at the start of the filter bank. An additional scaling is applied to account for variations in the absolute response among these KIDs, and to force this ratio to unity well away from the spectrometer channel passband. This ratio displays a minimum at the central frequency of the spectral channel, consistent with the expectation that the spectral channel is absorbing and reflecting a significant fraction of the incident power. A simple microwave circuit model shows that, on resonance, the scattering parameters of the optical circuit are $S_{21} = 1-Q_\mathrm{filt}/Q_\mathrm{feed}$, and $S_{11} = S_{21} - 1$. The ratio shown in the right panel of Figure~\ref{sockout_figure} is approximately $|S_{21}|^2$ of this circuit, with a correction for the contribution of a reflected component on the power detected by the first pair of broadband channels. Accounting for this reflection, the depth of this feature is determined by the ratio $Q_\mathrm{filter}/Q_\mathrm{feed}$, and the measured depth indicates $Q_\mathrm{filter}/Q_\mathrm{feed}\approx 0.25$. The channel was designed to have $Q_\mathrm{feed} = Q_\mathrm{det} = 500$, and for $Q_\mathrm{loss} = 1440$ we expect $Q_\mathrm{filter}/Q_\mathrm{feed}\approx 0.43$. This measurement then suggests that, even accounting for the $Q_\mathrm{loss}$ term, the channel is under-coupled.

On resonance, the fraction of the power on the feedline terminated in the detector is:

\begin{equation}
\eta_\mathrm{det} = 2\frac{Q_\mathrm{filt}}{Q_\mathrm{feed}}\bigg[1-\frac{Q_\mathrm{filt}}{Q_\mathrm{feed}}-\frac{Q_\mathrm{filt}}{Q_\mathrm{loss}}\bigg].
\end{equation}

\noindent Assuming $Q_\mathrm{loss} = 1440$, $Q_\mathrm{filter}/Q_\mathrm{feed} = 0.25$ yields $\eta_\mathrm{det} = 0.28$. This detection fraction may be increased for somewhat stronger coupling. For fixed $Q_\mathrm{filt}$ and $Q_\mathrm{loss}$, the detected power is maximized for $Q_\mathrm{filt}/Q_\mathrm{feed} = 0.5(1-Q_\mathrm{filt}/Q_\mathrm{loss})$, which yields 

\begin{equation}
\eta_\mathrm{det,max} = \frac{1}{2}\bigg[1-\frac{Q_\mathrm{filt}}{Q_\mathrm{loss}}\bigg]^2.
\end{equation}

\noindent For $Q_\mathrm{loss} = 1440$ this maximum detection efficiency is $0.34\rightarrow0.43$ for $Q_\mathrm{filt} = 250\rightarrow100$, achieved with $Q_\mathrm{filt}/Q_\mathrm{feed} = 0.41\rightarrow0.47$.

\subsection{Responsivity vs Optical Loading} \label{response_v_loading}

It is of interest to measure the dependence of the responsivity on optical loading, both from the standpoint of better understanding the properties of the TiN film, and to understand how best to engineer more sensitive devices. We have probed the response of our spectrometer channel to a large range of optical loading provided by our LO. The LO can produce up to $\sim$$2$ mW of power in a narrow band, significantly larger than the $\sim$$0.3$ pW of thermal power absorbed by the detector from the ambient 293 K background. However, the coupling efficiency between the spectrometer beam and the LO beam is low and difficult to estimate solely from the geometry of the optical setup. Calibrating the absolute power absorbed by the LO therefore requires additional measurement, and is ultimately grounded in observations of beam-filling thermal sources.

The LO contains a built-in, DC voltage-controlled RF attenuation. We have measured the attenuation as a function of applied voltage for several RF frequencies by coupling the LO to an Erickson Power Meter. The attenuation is repeatable, and attenuations of at least $\approx$$3$ dB can be applied without corrupting the final narrow-band signal (for attenuations $\gtrsim$$7$ dB the LO output can send power to multiple harmonics of the input RF tone). Use of this voltage-controlled attenuation then allows us to measure the response of our KID to fractional changes in the power coupled in from the LO, for fractional changes up to $\approx$$3$ dB.

\begin{figure}%
    \centering
    \subfloat[]{{\includegraphics[width=\myHalfFigureWidth\linewidth,keepaspectratio]{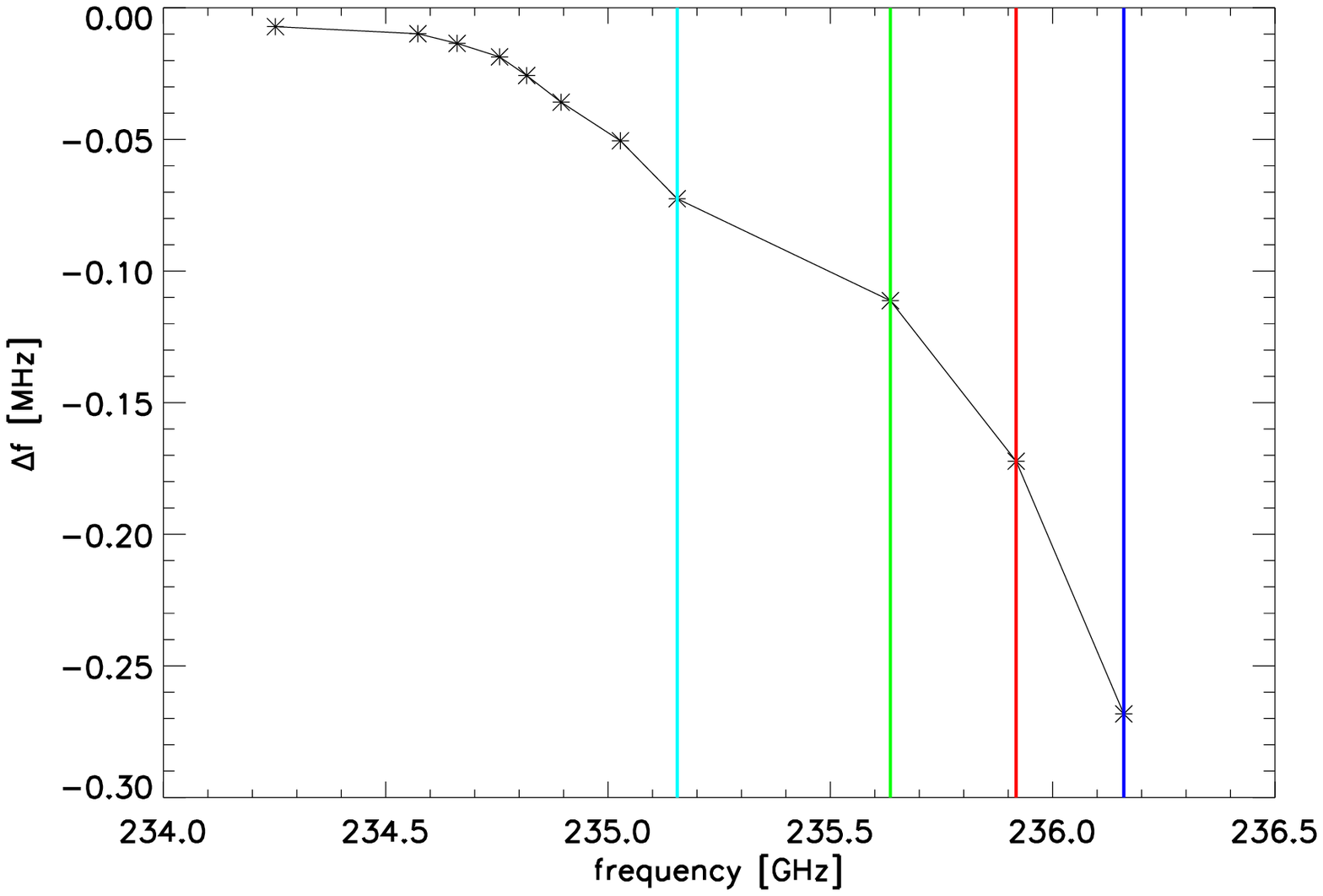} }}%
    \qquad
    \subfloat[]{{\includegraphics[width=\myHalfFigureWidth\linewidth,keepaspectratio]{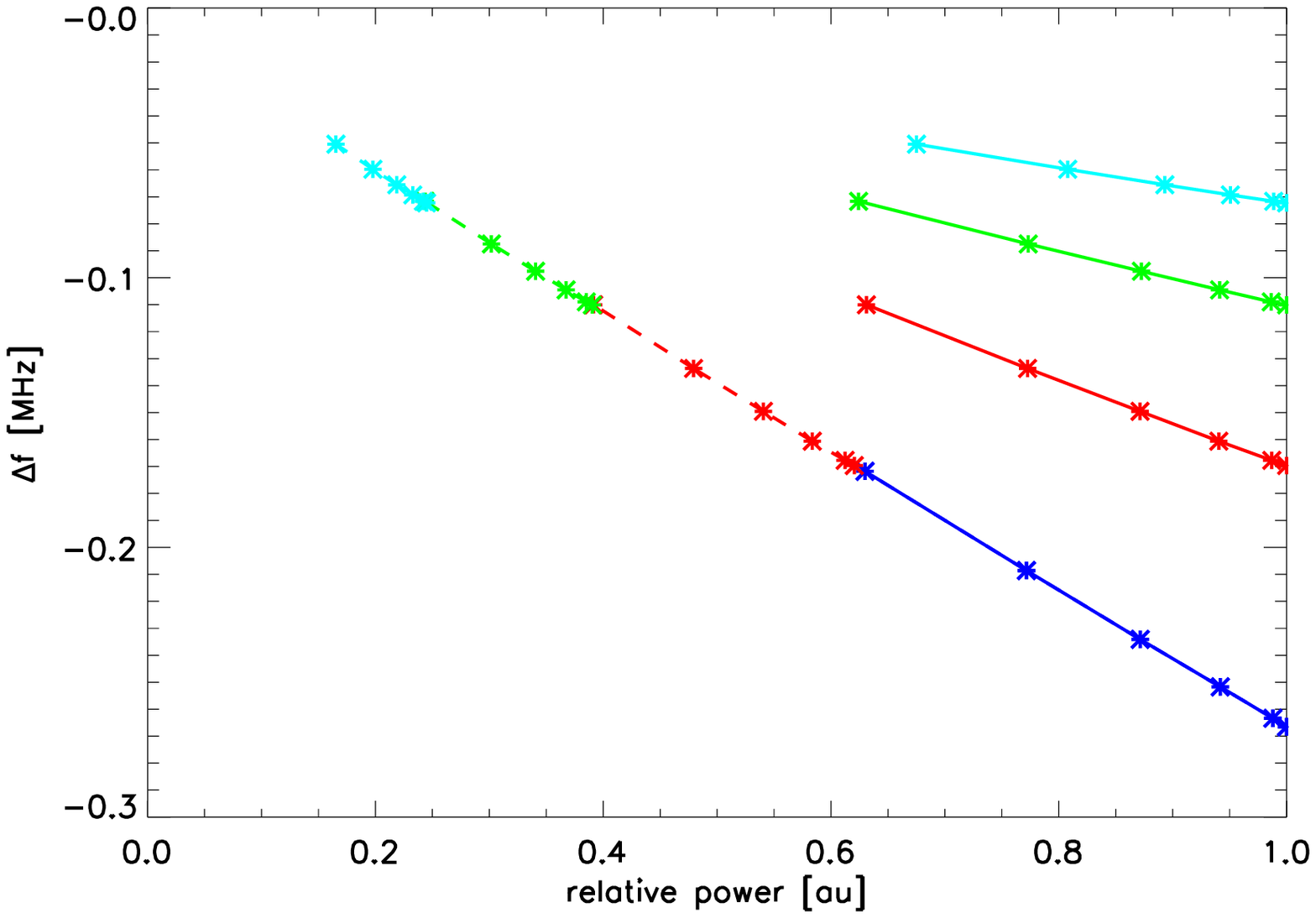} }}%
    \caption{(\textbf{a}) Frequency shift as a function of LO frequency for a single resonator, highlighting the peak response at $f_1 = 236.16$ GHz (\textit{blue}), and the response in the wings at $f_2 = 235.92$ GHz (\textit{red}), $f_3 = 235.63$ GHz (\textit{green}), and $f_4 = 235.16$ GHz (\textit{cyan}). (\textbf{b}) Resonator frequency shift for $f_1 - f_4$ as a function of relative LO power (\textit{solid lines}). Dashed lines show the results of scaling the curves for $f_2 - f_4$ to match and extend the $f_1$ curve for relative power $= 0.15 - 1$.}%
    \label{example_stitch}%
\end{figure}

To extend the dynamic range of this approach, we measure the response of the spectrometer with the LO operating at multiple frequencies distributed over the spectrometer passband. This approach is illustrated in Figure~\ref{example_stitch}. The left hand panel of Figure~\ref{example_stitch} shows the resonator frequency shift as a function of LO frequency, starting at the peak response at $f = 236.16$ GHz, and moving into the wings of the spectrometer response at lower frequencies. The response at four optical frequencies ($f_1 - f_4$) distributed over $\sim$$1$ GHz is highlighted. The right hand panel of Figure~\ref{example_stitch} shows the resonator frequency shift at each of these highlighted frequencies for various values of the voltage-controlled attenuation, from no attenuation to a maximum attenuation of $\approx$$2.2$ dB. These four optical frequencies were chosen such that resonator frequency shift with maximum attenuation for $f_1$ is the same as that for no attenuation at $f_2$, and a similar relation holds for the pairs $f_2/f_3$ and $f_3/f_4$. The plot of $\Delta f$ vs relative power for $f_2$ can then be scaled in the horizontal direction to match and extend the same curve for $f_1$. The curves for $f_3$ and $f_4$ can be similarly scaled, and stitching these four curves together yields a relation spanning 8 dB in relative power. This approach can be expanded to a yet larger range in optical loading by blocking a portion of the beam to reduce the loading by $3-6$ dB, and repeating this measurement. Repeating this procedure with three values of the bulk attenuation yields a measure of $\Delta f$ vs relative power spanning 14 dB in relative power.

\begin{figure}%
    \centering
    \subfloat[]{{\includegraphics[width=\myHalfFigureWidth\linewidth,keepaspectratio]{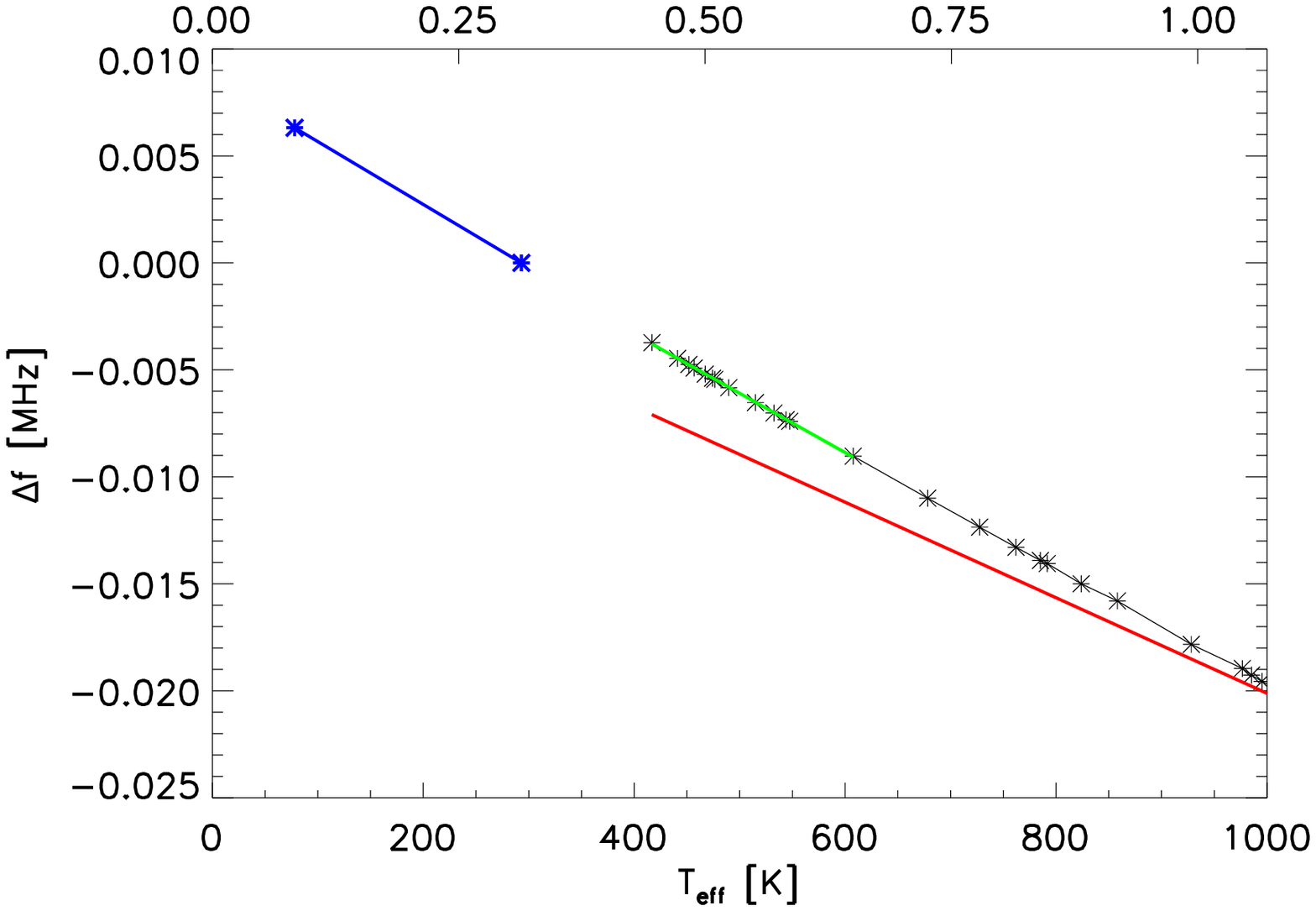} }}%
    \qquad
    \subfloat[]{{\includegraphics[width=\myHalfFigureWidth\linewidth,keepaspectratio]{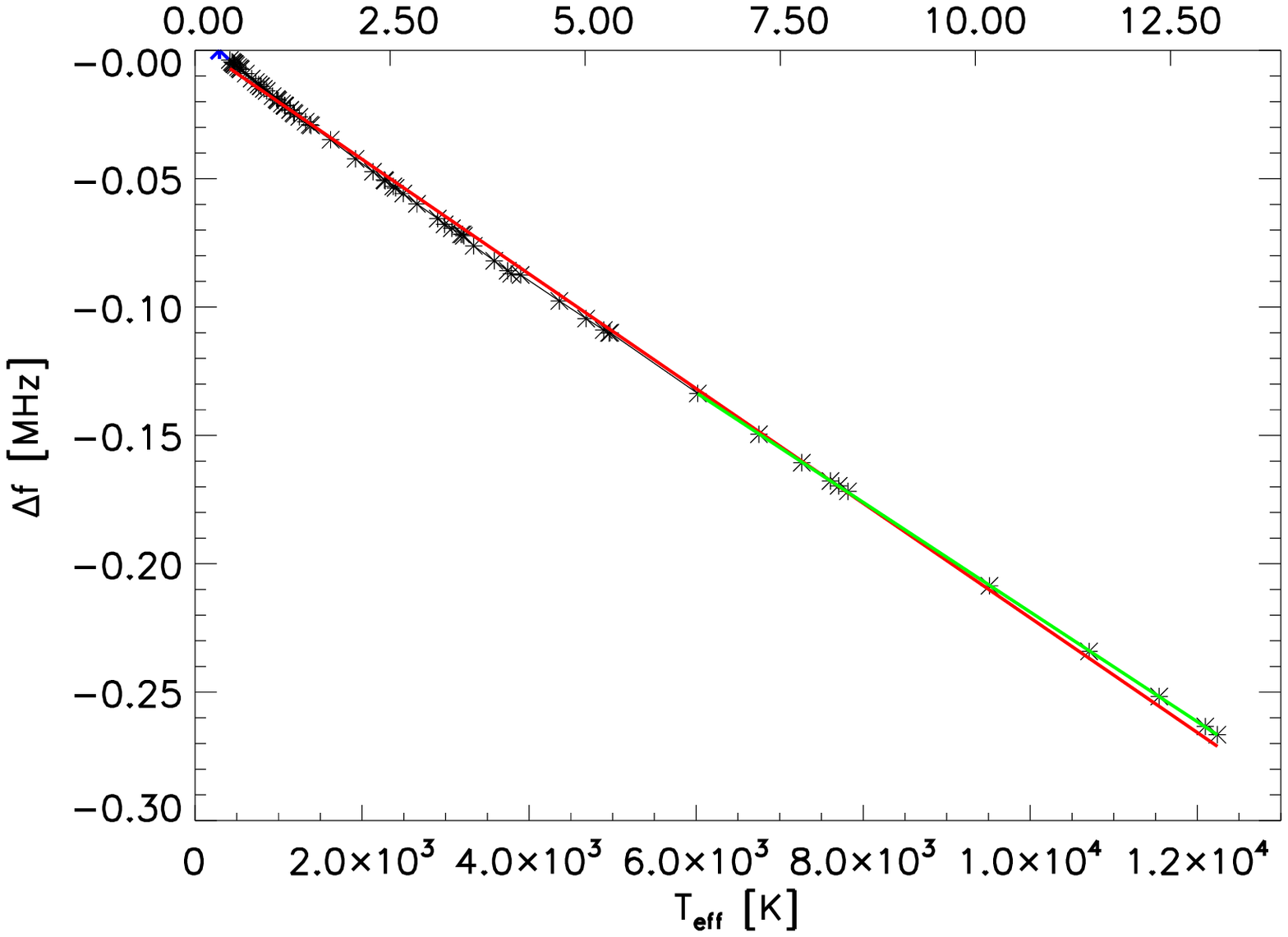} }}%
    \caption{(\textbf{a}) Resonator frequency shift $f_r - f_{r,0}$, where $f_{r,0}$ is the frequency with the cryostat exposed to beam-filling 293 K radiation. Black points show the frequency shift with added loading from the LO, and blue points show the shift when the ambient radiation temperature is reduced from 293 K to 78 K. Lower x-axis is the effective beam-filling radiation temperature at the front of the cryostat, and upper x-axis is the total loading at the detector. Red line is a linear fit to the full set of LO-controlled loading points shown in both panels, green line is a fit to the subset of the data at $\Delta f \ge -0.010$ MHz, or $T_\mathrm{eff} = 400 - 600$ K. (\textbf{b}) Same as panel a), with the green line a fit to the subset of the data at $\Delta f \le -0.12$ MHz, or $T_\mathrm{eff} = 6000 - 12000$ K.}%
    \label{fvload_figure}%
\end{figure}

This approach extends to a minimum frequency shift of $\approx$$4$ kHz. The frequency response to a beam-filling piece of AN-72 Eccosorb cooled from 293K to 78K is $\Delta f = 6.3$ kHz. We use this cold load calibration to assign an effective beam-filling radiation temperature to the LO measurements, by matching the slope of the $\Delta f/\Delta P$ relation for low LO loadings to that for the cold load. The result is shown in Figure~\ref{fvload_figure}. With this calibration, the power delivered by the LO corresponds to effective radiation temperatures $400 - 12000$ K, referenced to the cryostat window. A conversion from $T_\mathrm{eff}$ to power absorbed by the KID is shown for a system optical efficiency of $\eta_\mathrm{sys} = 0.059$ (see Section~\ref{noisesection}), and assuming no additional loading from within the cryostat.

%
%
%

The responsivity $\delta f/\delta P$ implied by the slope of the relation shown in Figure~\ref{fvload_figure} varies little with loading. Linear fits to subsets of the data at $T_\mathrm{eff} = 400 - 600$ K and $T_\mathrm{eff} = 6000 - 12000$ K yield slopes of $\Delta f/\Delta T_\mathrm{eff} = -28$ and $-21$ Hz/K, respectively, corresponding to $\delta f/\delta P \propto P^{-0.1}$ over the full range. In the limit that photon generated quasiparticles dominate the quasiparticle population, the quasiparticle density is expected to scale as the product of optical power and quasiparticle lifetime\cite{Zmuidzinas2012}: $n_\mathrm{qp}\propto\tau_\mathrm{qp}P$. For large optical loadings and hence large values of $n_\mathrm{qp}$, $\tau_\mathrm{qp} \propto 1/n_\mathrm{qp}$, and $n_\mathrm{qp}\propto P^{1/2}$. The resonator frequency shift is proportional to $\delta n_\mathrm{qp}$, and the responsivity then scales as $\delta f/\delta P \propto P^{-1/2}$. The relatively power-insensitive responsivity we find here runs counter to this model. However, similar power-independent responsivities for TiN KIDs have been reported elsewhere\cite{Hubmayr2014TiNphotonnoise}, suggesting the TiN response may require a modified model.

\subsection{Responsivity and System Efficiency} \label{noisesection}

\begin{figure}
\begin{center}
\includegraphics[%
  width=\myFigureWidth\linewidth,
  keepaspectratio]{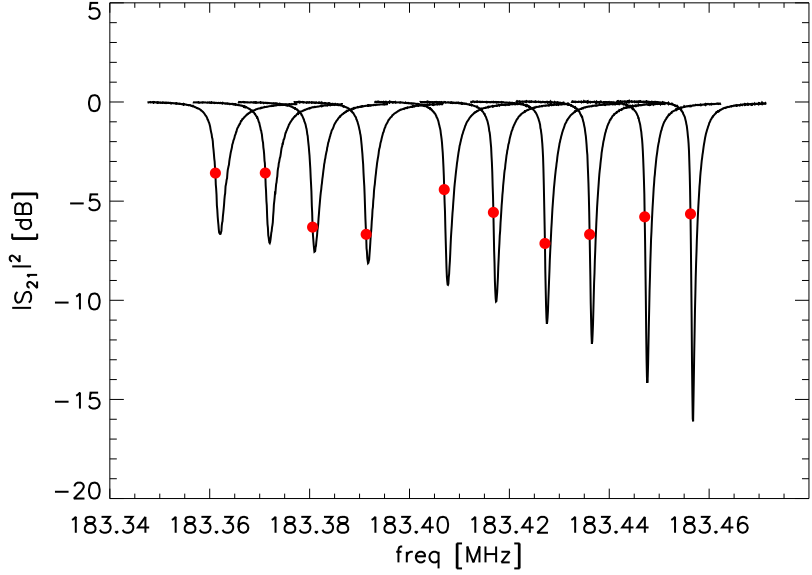}
\end{center}
\caption{Magnitude of the microwave transmission, $|S_\mathrm{21}|^2$, as a function of frequency for a range of optical loadings. From right to left the loading corresponds to an effective radiation temperature increasing from 293 K to 3540 K, when referenced to the cryostat entrance window. The solid points (\textit{red}) show the locations maximizing $\delta S_\mathrm{21}/\delta f$ at each loading, where the noise was measured.}
\label{mags21_figure}
\end{figure}

We estimate the intrinsic responsivity of our detectors, and the optical efficiency of the entire instrument, by measuring the scaling of photon noise with optical loading\cite{Yates2011,deVisser2014NatCo,Hubmayr2014TiNphotonnoise}. For each of a number of optical loadings, we begin by making a frequency sweep using a standard homodyne measurement (Figure~\ref{mags21_figure}). Fits to the complex transmission $S_\mathrm{21}(f)$ yield estimates of the resonant frequency $f_r$ and the resonator quality factor $Q_r$. We then obtain a noise trace at the frequency that maximizes $\delta S_\mathrm{21}/\delta f$, thereby minimizing the contribution of amplifier noise to the measured frequency noise. We project the raw $S_\mathrm{21}$ measurements into deviations in the frequency and dissipation directions, and compute the fractional frequency noise $S_\mathrm{xx}$. At bifurcation, the power on the line is $\approx$$-116$ dBm. We use a SiGe amplifier with a noise temperature of $T_n\approx6$ K, and even at bifurcation the amplifier noise makes a non-negligible contribution to the total frequency noise. To minimize this contribution, we make our noise measurements $\approx$$3$ dB below bifurcation. At this power level, the frequency maximizing $\delta S_\mathrm{21}/\delta f$ is lower than the resonant frequency $f_r$.

In operating so close to bifurcation we may be subject to non-linear effects in the KID, the most important of which is expected to be the power-dependent kinetic inductance. Specifically, reactive feedback has been shown to increase the fractional frequency noise in a TiN KID as the generator power is increased to the bifurcation level, and the KID is operated with negative detuning (as is done here)\cite{Swenson2013}. We have attempted to minimize any changes in the effects of nonlinearities on the device noise by increasing the generator power for measurements taken at higher optical loadings and lower $Q_r$, so as to keep the nonlinearity parameter $a$ approximately constant. In the measurements presented here, $Q_r$ decreases from 77,000 to 49,000 from lowest to highest loading, the generator power is increased by 4 dB over the same range, and $a$ varies non-monotonically between 0.28 and 0.49. Additionally, for each measurement we have verified that dropping the generator power by 3 dB results in only small changes to the measured device noise. This suggests the scaling of device noise with optical loading measured below is not strongly influenced by nonlinear effects in the KID.

\begin{figure}
\begin{center}
\includegraphics[%
  width=\myFigureWidth\linewidth,
  keepaspectratio]{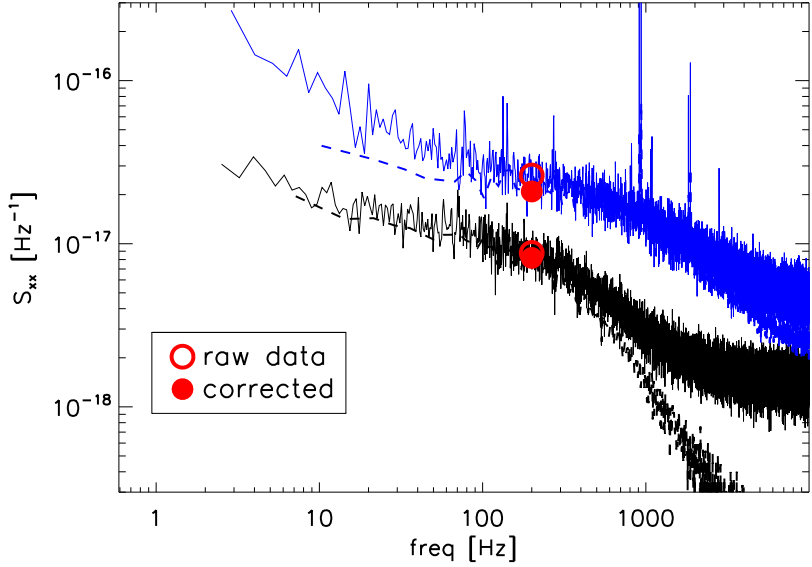}
\end{center}
\caption{Fractional frequency noise spectra for optical loadings corresponding to effective radiation temperatures of 293 K (\textit{black}) and 2500 K (\textit{blue}). Solid lines show the raw noise, and dashed lines show the estimated device noise, obtained by subtracting off the fitted contributions from the amplifier and 1/f components, and binning the residuals. Red points show the device noise estimated at 200 Hz by either averaging the raw noise over $f=150-250$ Hz (\textit{open points}), or using the results of a multiple component fit to isolate the intrinsic device noise (\textit{filled points}; see text).}
\label{psd_figure}
\end{figure}

The optical loading on the device is provided by nearly beam-filling Eccosorb at 293 K, with additional loading from our coherent source. In Figure~\ref{psd_figure} we show $S_\mathrm{xx}$ measured with the LO off, and with the LO tuned to provide an additional 2200 K loading. With the LO off, at $f \lesssim 200$ Hz $S_\mathrm{xx}$ is consistent with TLS noise with $S_\mathrm{xx,TLS} \propto f^{-0.25}$. This component rolls off at higher frequencies due to the finite resonator bandwidth, and at $f \gtrsim 3000$ Hz $S_\mathrm{xx}$ is dominated by white amplifier noise. With the LO turned on the device noise increases, and the resonator roll-off shifts to higher frequencies as $Q_r$ decreases. Additionally, a 1/f component becomes apparent. The amplitude of this component ($\sqrt{S_\mathrm{xx}}$) is approximately proportional to the optical loading, and measurement of the noise in multiple channels in parallel confirms it is highly correlated across the chip. The source of this 1/f component is likely drift in the LO power output, or in the transmission of some other element in the optical system.

We estimate the device noise at 200 Hz in two ways. First, we simply average the measured $S_\mathrm{xx}$ from $f = 150 - 250$ Hz. Second, we fit $S_\mathrm{xx}$ to a function of the form:

\begin{equation}
S_\mathrm{xx}(f) = \frac{Af^{-1} + Bf^{-p}}{1+(2\pi f\tau)^2} + C,
\end{equation}

\noindent where $Af^{-1}$ represents the 1/f component, $Bf^{-p}$ the device noise, $\tau$ parameterizes the roll-off, and $C$ is the amplifier noise. The intrinsic device noise at 200 Hz is then estimated as $S_\mathrm{xx,device} = B(200~\mathrm{Hz})^{-p}$. The estimated device noise, after subtracting the fitted amplifier and 1/f contributions, is shown in Figure~\ref{psd_figure}. There is tentative evidence that the spectral shape of the device noise flattens at higher optical loadings, as would be expected if the noise is becoming increasingly dominated by white photon and generation-recombination noise. However, the contamination from the 1/f component precludes a robust determination of the spectral slope. 

\begin{figure}
\begin{center} 
\includegraphics[%
  width=\myFigureWidth\linewidth,
  keepaspectratio]{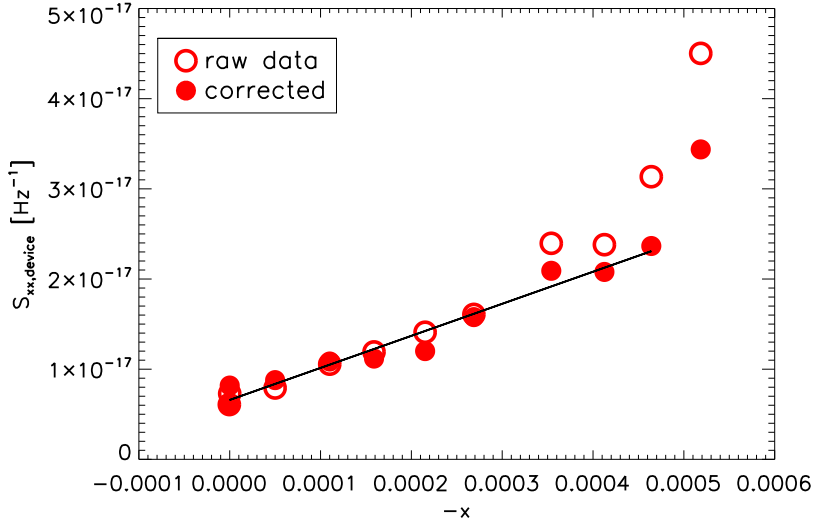}
\end{center}
\caption{Estimated device noise at 200 Hz against fractional frequency shift, obtained by either averaging the raw noise over $f=150-250$ Hz (\textit{open points}), or using the results of a multiple component fit to isolate the intrinsic device noise (\textit{filled points}). Two separate measurements are shown with the LO off ($x = 0$), obtained before and after the other measurements. Solid line shows a linear fit to the corrected points for $|x| \le 4.6\times10^{-4}$.}
\label{noisevloading_figure}
\end{figure}

In Figure~\ref{noisevloading_figure} we show the two estimates of the device noise at 200 Hz as a function of the fractional frequency shift $x = (f_r - f_{r,0})/f_{r,0}$, where $f_{r,0}$ is the resonator frequency with the LO off, and 293 K loading. The intrinsic device noise estimated through the multiple component decomposition of $S_\mathrm{xx}$ is consistent with having a linear dependence on $x$ for $|x| \lesssim 4.6\times10^{-4}$. For larger values of $|x|$ the contributions from the rising 1/f noise component can no longer be reliably subtracted. 
 
The total device noise is the sum of photon noise, generation-recombination noise, and TLS noise:

\begin{equation}
S_\mathrm{xx,device} = S_\mathrm{xx,LO} + S_\mathrm{xx,amb} + S_\mathrm{xx,GR} + S_\mathrm{xx,TLS},
\end{equation}

\noindent where the photon noise resulting from the loading provided by the LO is $S_\mathrm{xx,LO}$, and loading from the 293 K background and from within the cryostat produces a photon noise term $S_\mathrm{xx,amb}$. The photon noise produced by the LO emission is pure shot noise:

\begin{equation}
S_\mathrm{xx,LO} = R^2(2P_\mathrm{LO}h\nu),
\end{equation}

\noindent where $P_\mathrm{LO}$ is the absorbed power sourced by the LO, and $R = (dx/dP)$ is the fractional frequency responsivity for absorbed power $P$. Assuming the quasiparticle production is dominated by photon absorption, the generation-recombination noise can be written as\cite{deVisser2014NatCo}:

\begin{subequations}
\begin{align}
S_\mathrm{xx,GR} &= R^2\frac{2P\Delta}{\eta_\mathrm{pb}} \\
&=R^2\frac{2(P_\mathrm{LO} + P_\mathrm{amb})\Delta}{\eta_\mathrm{pb}},
\end{align}
\end{subequations}

\noindent where $P_\mathrm{amb}$ is the power received from all sources other than the LO, $\Delta$ is the superconducting energy gap, and $\eta_\mathrm{pb}$ is the pair-breaking efficiency. The TLS noise is expected to depend on the resonator internal power $P_\mathrm{int}$, and hence on the resonator quality factor $Q_r$ and the microwave generator power $P_g$ as\cite{Gao2008TLSmodel}:

\begin{subequations}
\begin{align}
S_\mathrm{xx,TLS} &\propto P_\mathrm{int}^{-1/2} \\
&\propto (Qr^2P_g)^{-1/2}.
\end{align}
\end{subequations}

\noindent For the measurements described here we increase the generator power at higher loadings and lower $Q_r$, and as a result $S_\mathrm{xx,TLS}$ is expected to vary by no more than $\pm 10\%$ over the full range of optical loading. Neglecting this variation, and noting that the photon noise and quasiparticle recombination noise associated with absorption of ambient power $P_\mathrm{amb}$ is independent of $P_\mathrm{LO}$, the change in device noise with fractional frequency shift can be written as:

\begin{subequations} \label{noise_equation}
\begin{align}
\frac{\partial S_\mathrm{xx,device}}{\partial x} &= \frac{\partial}{\partial x} \bigg[R^2(2P_\mathrm{LO}h\nu) + R^2\frac{2P_\mathrm{LO}\Delta}{\eta_\mathrm{pb}}\bigg] \\
&= R(2h\nu)\bigg[1+\frac{\Delta}{h\nu\eta_\mathrm{pb}}\bigg],
\end{align}
\end{subequations}

\noindent where we ignore any variation in responsivity with power. For an optical frequency of 236 GHz we have $h\nu/\Delta \approx 3.9$, and we use $\eta_\mathrm{pb} = 2\Delta/(h\nu)$ in Equation~\ref{noise_equation}, valid for $h\nu/\Delta \le 4$\cite{Guruswamy2014}.

The linear fit to $S_\mathrm{xx,device}$ shown in Figure~\ref{noisevloading_figure} is done for $|x| \le 4.6\times10^{-4}$, corresponding to an effective radiation temperature of $T_\mathrm{eff} = 293 - 3200$ K. For $R\propto P^{-0.1}$ the responsivity changes by $\pm10\%$ over this range, and the assumption of a power-independent responsivity used in Equation~\ref{noise_equation} is a reasonable approximation. Comparing the fitted slope with Equation~\ref{noise_equation} then yields a responsivity $R = -1.5\times10^8$\,W$^{-1}$. A calibrated cold load measurement then yields a full system optical efficiency of $\eta_\mathrm{sys} = 5.9\%$. Accounting for the finite transmission of the blocking and band-defining filters, the cold truncation of the edges of the antenna beam, Fresnel loss at the front of the bare silicon lens, and a 28\% spectrometer efficiency (Section~\ref{spectrometer_efficiency}), we expect a total system efficiency of 11\%. This is a factor of $\sim$$1.9$ larger than measured, indicating an additional source of loss in the system.

\section{Summary}

SuperSpec is an on-chip spectrometer we are developing for broadband, multi-object survey spectroscopy at submillimeter and millimeter wavelengths. We have presented the design and characterization of a new prototype device that delivers an improved sensitivity over the previous design. Our characterization of this device includes first measurements of the spectrometer efficiency, detector responsivity, and net optical efficiency of the full instrument.

For a representative $R=282$ channel centered at $f = 236$ GHz, we measure an instrument NEP of $5.2\times10^{-16}$ W\,Hz$^{-1/2}$, referenced to the front of the cryostat. This is a factor of $\sim$$10$ worse than the background-limited sensitivity of an $R=100$ spectrometer operating at the Caltech Submillimeter Observatory, with a single polarization system transmission of 25\%. However, we have identified a number of ways to improve our current sensitivity with only moderate changes to the spectrometer design. The optical efficiency of the system may be increased by applying an anti-reflection coating to the silicon lens, optimizing the coupling factors $Q_\mathrm{feed}$ and $Q_\mathrm{det}$ to maximize the spectrometer detection efficiency, and identifying and removing the source of the additional factor of $\sim$$1.9$ reduction in system transmission (Section~\ref{noisesection}). We will increase the responsivity in future devices by reducing the TiN indutor volume, and decreasing the $T_c$ from 1.65 K to 1.2 K. We are also exploring options to reduce the TLS noise with modifications to the layout of the KID IDC. The net sensitivity gain from these changes is expected to provide the requisite factor of $\sim$$10$ reduction in NEP that will enable the construction of an $R=100$ background-limited filter bank spectrometer. This resolving power is sufficiently high for a line intensity mapping experiment targeting [CII] 158 $\mu$m emission during the Epoch of Reionization.

\acknowledgments     
 
This project is supported by NASA Astrophysics Research and Analysis (APRA) grant no. 399131.02.06.03.43. Part of this research was carried out at the Jet Propulsion Laboratory, California Institute of Technology, under a contract with the National Aeronautics and Space Administration.

\bibliography{List}

\begin{thebibliography}{10}

\bibitem{Fixsen1998}
{Fixsen}, D.~J., {Dwek}, E., {Mather}, J.~C., {Bennett}, C.~L., and {Shafer},
  R.~A., ``{The Spectrum of the Extragalactic Far-Infrared Background from the
  COBE FIRAS Observations},'' {\em \apj}~{\bf 508},  123--128 (Nov. 1998).

\bibitem{Berta2010}
{Berta}, S., {Magnelli}, B., {Lutz}, D., {Altieri}, B., {Aussel}, H.,
  {Andreani}, P., {Bauer}, O., {Bongiovanni}, A., {Cava}, A., {Cepa}, J.,
  {Cimatti}, A., {Daddi}, E., {Dominguez}, H., {Elbaz}, D., {Feuchtgruber}, H.,
  {F{\"o}rster Schreiber}, N.~M., {Genzel}, R., {Gruppioni}, C., {Katterloher},
  R., {Magdis}, G., {Maiolino}, R., {Nordon}, R., {P{\'e}rez Garc{\'{\i}}a},
  A.~M., {Poglitsch}, A., {Popesso}, P., {Pozzi}, F., {Riguccini}, L.,
  {Rodighiero}, G., {Saintonge}, A., {Santini}, P., {Sanchez-Portal}, M.,
  {Shao}, L., {Sturm}, E., {Tacconi}, L.~J., {Valtchanov}, I., {Wetzstein}, M.,
  and {Wieprecht}, E., ``{Dissecting the cosmic infra-red background with
  Herschel/PEP},'' {\em \aap}~{\bf 518},  L30 (July 2010).

\bibitem{Vieira2013}
{Vieira}, J.~D., {Marrone}, D.~P., {Chapman}, S.~C., {De Breuck}, C.,
  {Hezaveh}, Y.~D., {Wei{$\beta$}}, A., {Aguirre}, J.~E., {Aird}, K.~A.,
  {Aravena}, M., {Ashby}, M.~L.~N., {Bayliss}, M., {Benson}, B.~A., {Biggs},
  A.~D., {Bleem}, L.~E., {Bock}, J.~J., {Bothwell}, M., {Bradford}, C.~M.,
  {Brodwin}, M., {Carlstrom}, J.~E., {Chang}, C.~L., {Crawford}, T.~M.,
  {Crites}, A.~T., {de Haan}, T., {Dobbs}, M.~A., {Fomalont}, E.~B.,
  {Fassnacht}, C.~D., {George}, E.~M., {Gladders}, M.~D., {Gonzalez}, A.~H.,
  {Greve}, T.~R., {Gullberg}, B., {Halverson}, N.~W., {High}, F.~W., {Holder},
  G.~P., {Holzapfel}, W.~L., {Hoover}, S., {Hrubes}, J.~D., {Hunter}, T.~R.,
  {Keisler}, R., {Lee}, A.~T., {Leitch}, E.~M., {Lueker}, M., {Luong-van}, D.,
  {Malkan}, M., {McIntyre}, V., {McMahon}, J.~J., {Mehl}, J., {Menten}, K.~M.,
  {Meyer}, S.~S., {Mocanu}, L.~M., {Murphy}, E.~J., {Natoli}, T., {Padin}, S.,
  {Plagge}, T., {Reichardt}, C.~L., {Rest}, A., {Ruel}, J., {Ruhl}, J.~E.,
  {Sharon}, K., {Schaffer}, K.~K., {Shaw}, L., {Shirokoff}, E., {Spilker},
  J.~S., {Stalder}, B., {Staniszewski}, Z., {Stark}, A.~A., {Story}, K.,
  {Vanderlinde}, K., {Welikala}, N., and {Williamson}, R., ``{Dusty starburst
  galaxies in the early Universe as revealed by gravitational lensing},'' {\em
  \nat}~{\bf 495},  344--347 (Mar. 2013).

\bibitem{Geach2013SCUBA2}
{Geach}, J.~E., {Chapin}, E.~L., {Coppin}, K.~E.~K., {Dunlop}, J.~S.,
  {Halpern}, M., {Smail}, I., {Werf}, P.~v.~d., {Serjeant}, S., {Farrah}, D.,
  {Roseboom}, I., {Targett}, T., {Arumugam}, V., {Asboth}, V., {Blain}, A.,
  {Chrysostomou}, A., {Clarke}, C., {Ivison}, R.~J., {Jones}, S.~L., {Karim},
  A., {Mackenzie}, T., {Meijerink}, R., {Micha{\l}owski}, M.~J., {Scott}, D.,
  {Simpson}, J.~M., {Swinbank}, A.~M., {Alexander}, D.~M., {Almaini}, O.,
  {Aretxaga}, I., {Best}, P., {Chapman}, S., {Clements}, D.~L., {Conselice},
  C., {Danielson}, A.~L.~R., {Eales}, S., {Edge}, A.~C., {Gibb}, A.~G.,
  {Hughes}, D., {Jenness}, T., {Knudsen}, K.~K., {Lacey}, C.~G., {Marsden}, G.,
  {McMahon}, R., {Oliver}, S.~J., {Page}, M.~J., {Peacock}, J.~A.,
  {Rigopoulou}, D., {Robson}, E.~I., {Spaans}, M., {Stevens}, J., {Webb},
  T.~M.~A., {Willott}, C., {Wilson}, C.~D., and {Zemcov}, M., ``{The SCUBA-2
  Cosmology Legacy Survey: blank-field number counts of 450-{$\mu$}m-selected
  galaxies and their contribution to the cosmic infrared background},'' {\em
  \mnras}~{\bf 432},  53--61 (June 2013).

\bibitem{Dowell2014}
{Dowell}, C.~D., {Conley}, A., {Glenn}, J., {Arumugam}, V., {Asboth}, V.,
  {Aussel}, H., {Bertoldi}, F., {B{\'e}thermin}, M., {Bock}, J., {Boselli}, A.,
  {Bridge}, C., {Buat}, V., {Burgarella}, D., {Cabrera-Lavers}, A., {Casey},
  C.~M., {Chapman}, S.~C., {Clements}, D.~L., {Conversi}, L., {Cooray}, A.,
  {Dannerbauer}, H., {De Bernardis}, F., {Ellsworth-Bowers}, T.~P., {Farrah},
  D., {Franceschini}, A., {Griffin}, M., {Gurwell}, M.~A., {Halpern}, M.,
  {Hatziminaoglou}, E., {Heinis}, S., {Ibar}, E., {Ivison}, R.~J., {Laporte},
  N., {Marchetti}, L., {Mart{\'{\i}}nez-Navajas}, P., {Marsden}, G.,
  {Morrison}, G.~E., {Nguyen}, H.~T., {O'Halloran}, B., {Oliver}, S.~J.,
  {Omont}, A., {Page}, M.~J., {Papageorgiou}, A., {Pearson}, C.~P., {Petitpas},
  G., {P{\'e}rez-Fournon}, I., {Pohlen}, M., {Riechers}, D., {Rigopoulou}, D.,
  {Roseboom}, I.~G., {Rowan-Robinson}, M., {Sayers}, J., {Schulz}, B., {Scott},
  D., {Seymour}, N., {Shupe}, D.~L., {Smith}, A.~J., {Streblyanska}, A.,
  {Symeonidis}, M., {Vaccari}, M., {Valtchanov}, I., {Vieira}, J.~D., {Viero},
  M., {Wang}, L., {Wardlow}, J., {Xu}, C.~K., and {Zemcov}, M., ``{HerMES:
  Candidate High-redshift Galaxies Discovered with Herschel/SPIRE},'' {\em
  \apj}~{\bf 780},  75 (Jan. 2014).

\bibitem{Stacey2010CII}
{Stacey}, G.~J., {Hailey-Dunsheath}, S., {Ferkinhoff}, C., {Nikola}, T.,
  {Parshley}, S.~C., {Benford}, D.~J., {Staguhn}, J.~G., and {Fiolet}, N., ``{A
  158 {$\mu$}m [C II] Line Survey of Galaxies at z \~{} 1-2: An Indicator of
  Star Formation in the Early Universe},'' {\em \apj}~{\bf 724},  957--974
  (Dec. 2010).

\bibitem{Riechers2013Nature}
{Riechers}, D.~A., {Bradford}, C.~M., {Clements}, D.~L., {Dowell}, C.~D.,
  {P{\'e}rez-Fournon}, I., {Ivison}, R.~J., {Bridge}, C., {Conley}, A., {Fu},
  H., {Vieira}, J.~D., {Wardlow}, J., {Calanog}, J., {Cooray}, A., {Hurley},
  P., {Neri}, R., {Kamenetzky}, J., {Aguirre}, J.~E., {Altieri}, B.,
  {Arumugam}, V., {Benford}, D.~J., {B{\'e}thermin}, M., {Bock}, J.,
  {Burgarella}, D., {Cabrera-Lavers}, A., {Chapman}, S.~C., {Cox}, P.,
  {Dunlop}, J.~S., {Earle}, L., {Farrah}, D., {Ferrero}, P., {Franceschini},
  A., {Gavazzi}, R., {Glenn}, J., {Solares}, E.~A.~G., {Gurwell}, M.~A.,
  {Halpern}, M., {Hatziminaoglou}, E., {Hyde}, A., {Ibar}, E., {Kov{\'a}cs},
  A., {Krips}, M., {Lupu}, R.~E., {Maloney}, P.~R., {Martinez-Navajas}, P.,
  {Matsuhara}, H., {Murphy}, E.~J., {Naylor}, B.~J., {Nguyen}, H.~T., {Oliver},
  S.~J., {Omont}, A., {Page}, M.~J., {Petitpas}, G., {Rangwala}, N.,
  {Roseboom}, I.~G., {Scott}, D., {Smith}, A.~J., {Staguhn}, J.~G.,
  {Streblyanska}, A., {Thomson}, A.~P., {Valtchanov}, I., {Viero}, M., {Wang},
  L., {Zemcov}, M., and {Zmuidzinas}, J., ``{A dust-obscured massive
  maximum-starburst galaxy at a redshift of 6.34},'' {\em \nat}~{\bf 496},
  329--333 (Apr. 2013).

\bibitem{Weiss2009}
{Wei{\ss}}, A., {Ivison}, R.~J., {Downes}, D., {Walter}, F., {Cirasuolo}, M.,
  and {Menten}, K.~M., ``{First Redshift Determination of an
  Optically/Ultraviolet Faint Submillimeter Galaxy Using CO Emission Lines},''
  {\em \apjl}~{\bf 705},  L45--L47 (Nov. 2009).

\bibitem{Walter2012Nature}
{Walter}, F., {Decarli}, R., {Carilli}, C., {Bertoldi}, F., {Cox}, P., {da
  Cunha}, E., {Daddi}, E., {Dickinson}, M., {Downes}, D., {Elbaz}, D., {Ellis},
  R., {Hodge}, J., {Neri}, R., {Riechers}, D.~A., {Weiss}, A., {Bell}, E.,
  {Dannerbauer}, H., {Krips}, M., {Krumholz}, M., {Lentati}, L., {Maiolino},
  R., {Menten}, K., {Rix}, H.-W., {Robertson}, B., {Spinrad}, H., {Stark},
  D.~P., and {Stern}, D., ``{The intense starburst HDF~850.1 in a galaxy
  overdensity at z$\sim$5.2 in the Hubble Deep Field},'' {\em \nat}~{\bf 486},
  233--236 (June 2012).

\bibitem{Gong2012}
{Gong}, Y., {Cooray}, A., {Silva}, M., {Santos}, M.~G., {Bock}, J., {Bradford},
  C.~M., and {Zemcov}, M., ``{Intensity Mapping of the [C II] Fine Structure
  Line during the Epoch of Reionization},'' {\em \apj}~{\bf 745},  49 (Jan.
  2012).

\bibitem{Salvaterra2011}
{Salvaterra}, R., {Ferrara}, A., and {Dayal}, P., ``{Simulating high-redshift
  galaxies},'' {\em \mnras}~{\bf 414},  847--859 (June 2011).

\bibitem{Kovacs2012SPIE}
{Kov{\'a}cs}, A., {Barry}, P.~S., {Bradford}, C.~M., {Chattopadhyay}, G.,
  {Day}, P., {Doyle}, S., {Hailey-Dunsheath}, S., {Hollister}, M., {McKenney},
  C., {LeDuc}, H.~G., {Llombart}, N., {Marrone}, D.~P., {Mauskopf}, P.,
  {O'Brient}, R.~C., {Padin}, S., {Swenson}, L.~J., and {Zmuidzinas}, J.,
  ``{SuperSpec: design concept and circuit simulations},'' in [{\em Society of
  Photo-Optical Instrumentation Engineers (SPIE) Conference
  Series}{\nolinebreak\hspace{0.1em}]},  {\em Society of Photo-Optical
  Instrumentation Engineers (SPIE) Conference Series} {\bf 8452} (Sept. 2012).

\bibitem{Shirokoff2012SPIE}
{Shirokoff}, E., {Barry}, P.~S., {Bradford}, C.~M., {Chattopadhyay}, G., {Day},
  P., {Doyle}, S., {Hailey-Dunsheath}, S., {Hollister}, M.~I., {Kov{\'a}cs},
  A., {McKenney}, C., {Leduc}, H.~G., {Llombart}, N., {Marrone}, D.~P.,
  {Mauskopf}, P., {O'Brient}, R., {Padin}, S., {Reck}, T., {Swenson}, L.~J.,
  and {Zmuidzinas}, J., ``{MKID development for SuperSpec: an on-chip, mm-wave,
  filter-bank spectrometer},'' in [{\em Society of Photo-Optical
  Instrumentation Engineers (SPIE) Conference
  Series}{\nolinebreak\hspace{0.1em}]},  {\em Society of Photo-Optical
  Instrumentation Engineers (SPIE) Conference Series} {\bf 8452} (Sept. 2012).

\bibitem{Barry2012SPIE}
{Barry}, P.~S., {Shirokoff}, E., {Kov{\'a}cs}, A., {Reck}, T.~J.,
  {Hailey-Dunsheath}, S., {McKenney}, C.~M., {Swenson}, L.~J., {Hollister},
  M.~I., {Leduc}, H.~G., {Doyle}, S., {O'Brient}, R., {Llombart}, N.,
  {Marrone}, D., {Chattopadhyay}, G., {Day}, P.~K., {Padin}, S., {Bradford},
  C.~M., {Mauskopf}, P.~D., and {Zmuidzinas}, J., ``{Electromagnetic design for
  SuperSpec: a lithographically-patterned millimetre-wave spectrograph},'' in
  [{\em Society of Photo-Optical Instrumentation Engineers (SPIE) Conference
  Series}{\nolinebreak\hspace{0.1em}]},  {\em Society of Photo-Optical
  Instrumentation Engineers (SPIE) Conference Series} {\bf 8452} (Sept. 2012).

\bibitem{LeDuc2010}
{Leduc}, H.~G., {Bumble}, B., {Day}, P.~K., {Eom}, B.~H., {Gao}, J., {Golwala},
  S., {Mazin}, B.~A., {McHugh}, S., {Merrill}, A., {Moore}, D.~C., {Noroozian},
  O., {Turner}, A.~D., and {Zmuidzinas}, J., ``{Titanium nitride films for
  ultrasensitive microresonator detectors},'' {\em Applied Physics
  Letters}~{\bf 97},  102509 (Sept. 2010).

\bibitem{Zmuidzinas2012}
Zmuidzinas, J., ``Superconducting microresonators: Physics and applications,''
  {\em Annual Review of Condensed Matter Physics}~{\bf 3}(1),  169--214 (2012).

\bibitem{HaileyDunsheath2014LTD}
{Hailey-Dunsheath}, S., {Barry}, P.~S., {Bradford}, C.~M., {Chattopadhyay}, G.,
  {Day}, P., {Doyle}, S., {Hollister}, M., {Kovacs}, A., {LeDuc}, H.~G.,
  {Llombart}, N., {Mauskopf}, P., {McKenney}, C., {Monroe}, R., {Nguyen},
  H.~T., {O'Brient}, R., {Padin}, S., {Reck}, T., {Shirokoff}, E., {Swenson},
  L., {Tucker}, C.~E., and {Zmuidzinas}, J., ``{Optical Measurements of
  SuperSpec: A Millimeter-Wave On-Chip Spectrometer},'' {\em Journal of Low
  Temperature Physics}  (Jan. 2014).

\bibitem{Shirokoff2014LTD}
{Shirokoff}, E., {Barry}, P.~S., {Bradford}, C.~M., {Chattopadhyay}, G., {Day},
  P., {Doyle}, S., {Hailey-Dunsheath}, S., {Hollister}, M.~I., {Kov{\'a}cs},
  A., {Leduc}, H.~G., {McKenney}, C.~M., {Mauskopf}, P., {Nguyen}, H.~T.,
  {O'Brient}, R., {Padin}, S., {Reck}, T.~J., {Swenson}, L.~J., {Tucker},
  C.~E., and {Zmuidzinas}, J., ``{Design and Performance of SuperSpec: An
  On-Chip, KID-Based, mm-Wavelength Spectrometer},'' {\em Journal of Low
  Temperature Physics}  (Feb. 2014).

\bibitem{Leech2011}
{Leech}, J., {Tan}, B.~K., {Yassin}, G., {Kittara}, P., {Wangsuya}, S.,
  {Treuttel}, J., {Henry}, M., {Oldfield}, M.~L., and {Huggard}, P.~G.,
  ``{Multiple flare-angle horn feeds for sub-mm astronomy and cosmic microwave
  background experiments},'' {\em \aap}~{\bf 532},  A61 (Aug. 2011).

\bibitem{McKenney2012}
{McKenney}, C.~M., {Leduc}, H.~G., {Swenson}, L.~J., {Day}, P.~K., {Eom},
  B.~H., and {Zmuidzinas}, J., ``{Design considerations for a background
  limited 350 micron pixel array using lumped element superconducting
  microresonators},'' in [{\em Society of Photo-Optical Instrumentation
  Engineers (SPIE) Conference Series}{\nolinebreak\hspace{0.1em}]},  {\em
  Society of Photo-Optical Instrumentation Engineers (SPIE) Conference Series}
  {\bf 8452} (Sept. 2012).

\bibitem{Filipovic1993ITMTT}
{Filipovic}, D.~F., {Gearhart}, S.~S., and {Rebeiz}, G.~M., ``{Double-slot
  antennas on extended hemispherical and elliptical silicon dielectric
  lenses},'' {\em IEEE Transactions on Microwave Theory Techniques}~{\bf 41},
  1738--1749 (Oct. 1993).

\bibitem{Hubmayr2014TiNphotonnoise}
{Hubmayr}, J., {Beall}, J., {Becker}, D., {Cho}, H.-M., {Devlin}, M., {Dober},
  B., {Groppi}, C., {Hilton}, G.~C., {Irwin}, K.~D., {Li}, D., {Mauskopf}, P.,
  {Pappas}, D.~P., {Van Lanen}, J., {Vissers}, M.~R., and {Gao}, J.,
  ``{Photon-noise limited sensitivity in titanium nitride kinetic inductance
  detectors},'' {\em ArXiv e-prints}  (June 2014).

\bibitem{Yates2011}
Yates, S. J.~C., Baselmans, J. J.~A., Endo, A., Janssen, R. M.~J., Ferrari, L.,
  Diener, P., and Baryshev, A.~M., ``Photon noise limited radiation detection
  with lens-antenna coupled microwave kinetic inductance detectors,'' {\em
  Applied Physics Letters}~{\bf 99}(7),  -- (2011).

\bibitem{deVisser2014NatCo}
{de Visser}, P.~J., {Baselmans}, J.~J.~A., {Bueno}, J., {Llombart}, N., and
  {Klapwijk}, T.~M., ``{Fluctuations in the electron system of a superconductor
  exposed to a photon flux},'' {\em Nature Communications}~{\bf 5} (Feb. 2014).

\bibitem{Swenson2013}
{Swenson}, L.~J., {Day}, P.~K., {Eom}, B.~H., {Leduc}, H.~G., {Llombart}, N.,
  {McKenney}, C.~M., {Noroozian}, O., and {Zmuidzinas}, J., ``{Operation of a
  titanium nitride superconducting microresonator detector in the nonlinear
  regime},'' {\em Journal of Applied Physics}~{\bf 113},  104501 (Mar. 2013).

\bibitem{Gao2008TLSmodel}
{Gao}, J., {Daal}, M., {Martinis}, J.~M., {Vayonakis}, A., {Zmuidzinas}, J.,
  {Sadoulet}, B., {Mazin}, B.~A., {Day}, P.~K., and {Leduc}, H.~G., ``{A
  semiempirical model for two-level system noise in superconducting
  microresonators},'' {\em Applied Physics Letters}~{\bf 92},  212504 (May
  2008).

\bibitem{Guruswamy2014}
{Guruswamy}, T., {Goldie}, D.~J., and {Withington}, S., ``{Quasiparticle
  generation efficiency in superconducting thin films},'' {\em Superconductor
  Science Technology}~{\bf 27},  055012 (May 2014).

\end{thebibliography}
\bibliographystyle{spiebib}
\end{document}